\documentclass{emulateapj}
\usepackage{natbib}
\usepackage{epsfig} 
\usepackage{graphicx}
\usepackage{subfigure}
\usepackage{float}
\usepackage{amsmath}
\usepackage{color}
\usepackage{amssymb} 
\usepackage{amsfonts}
\usepackage{units}
\usepackage[colorlinks,linkcolor=blue,anchorcolor=green,citecolor=blue]{hyperref}
\shorttitle{Large host-galaxy DM of FRBs} 
\shortauthors{Yang et al.}
\begin{document} 

\title{Large host-galaxy dispersion measure of fast radio bursts}

\author{Yuan-Pei Yang\altaffilmark{1}, Rui Luo\altaffilmark{1,2}, Zhuo Li\altaffilmark{1,2} and Bing Zhang\altaffilmark{1,2,3}}

\affil{$^1$Kavli Institute for Astronomy and Astrophysics, Peking University, Beijing 100871, China; KIAA-CAS Fellow, yypspore@gmail.com; luorui1991@pku.edu.cn;  zhuo.li@pku.edu.cn;\\
$^2$ Department of Astronomy, School of Physics, Peking University, Beijing 100871, China \\
$^3$ Department of Physics and Astronomy, University of Nevada, Las Vegas, NV 89154, USA; zhang@physics.unlv.edu}

\begin{abstract}
Fast radio bursts (FRBs) have excessive dispersion measures (DMs) and an all-sky distribution, which point toward an extragalactic or even a cosmological origin. We develop a method to extract the mean host galaxy DM ($\left\langle{\rm DM_{HG,loc}}\right\rangle$) and the characterized luminosity ($L$) of FRBs using the observed DM-Flux data, based on the assumption of a narrow luminosity distribution. Applying Bayesian inference to the data of 21 FRBs, we derive a relatively large mean host DM, i.e. $\left\langle{\rm DM_{HG,loc}}\right\rangle \sim 270~\unit{pc~cm^{-3}}$ with a large dispersion. A relatively large ${\rm DM_{HG}}$ of FRBs is also supported by the millisecond scattering times of some FRBs and the relatively small redshift $z=0.19273$ of FRB 121102 (which gives ${\rm DM_{HG,loc}} \sim 210 ~\unit{pc~cm^{-3}}$). The large host galaxy DM may be contributed by the ISM or a near-source plasma in the host galaxy. If it is contributed by the ISM, the type of the FRB host galaxies would not be Milky Way (MW)-like, consistent with the detected host of FRB 121102. We also discuss the possibility of having a near-source supernova remnant (SNR), pulsar wind nebula (PWN) or HII region that gives a significant contribution to the observed ${\rm DM_{HG}}$. 
\end{abstract}


\section{Introduction}

Fast radio bursts (FRBs) are mysterious astronomical radio transients with short intrinsic durations ($\sim1\unit{ms}$), large dispersion measures (${\rm DM}\gtrsim200~\unit{pc~cm^{-3}}$), and an all-sky distribution \citep{lor07,kea12,kea16,tho13,bur14,spi14,spi16,mas15,pet15,rav15,rav16,cha16,cha17,cal17}. Recently, thanks to the precise localization and multi-wavelength follow-up observations of the repeating source FRB 121102 \citep{cha17,mar17,ten17}, the distance scale of FRBs has been finally settled to a cosmological scale at $z=0.19273$ \citep{ten17}. The large DM excess of other FRBs with respect to the Galactic value and their high Galactic latitudes also suggest that most, if not all, FRBs should have an extragalactic (e.g. $\sim 10$ Mpc), and likely a cosmological (e.g. $> 100$ Mpc) origin.

The host galaxies of FRBs carry important information regarding the progenitor of FRBs. For FRB 121102, optical imaging and spectroscopy indicate a dwarf galaxy with a mass of $M\sim(4-7)\times10^7M_\odot$ as the host galaxy. The $\rm{H_\alpha}$ flux of the host galaxy suggests a star formation rate of ${\rm SFR}\sim0.4M_\odot \unit{yr^{-1}}$ \citep{ten17}. No information about the host galaxies of other FRBs is available.
One possible way to derive FRB host galaxy information is to extract the host galaxy DM from data.   
\cite{yan16} proposed a method to derive ${\rm DM_{HG}}$ using the measured DM and $z$ of a sample of FRBs. However, the $z$ values of most FRBs are not obtained so far.

In this paper, we further develop a method to apply DM and flux of FRBs to infer ${\rm DM_{HG}}$. This method is applied to the current FRB sample with 21 sources. 
Through Bayesian inference, we derive a relatively large mean host galaxy DM, $\left\langle{\rm DM_{HG,loc}}\right\rangle$, for FRBs. We also provide two pieces of supporting evidence for a large value of ${\rm DM_{HG}}$: millisecond-duration of scattering tails for some FRBs and ${\rm DM_{HG,loc}} \sim 210 ~\unit{pc~cm^{-3}}$ for FRB 121102.

\section{Method}

For an FRB, the observed dispersion measure has three contributions \citep[e.g.][]{lor07,kea12,tho13,den14,gao14,yan16,mur16a}, i.e.
\begin{eqnarray}
{\rm DM_{obs}}={\rm DM_{MW}}+{\rm DM_{HG}}+{\rm DM_{IGM}},
\end{eqnarray}
which are from the Milky Way, the FRB host galaxy (which itself includes the contributions from the interstellar medium (ISM) in the host galaxy and  a near-source plasma), and the IGM, respectively. According to the Galactic pulsar data, ${\rm DM_{MW}}$ can be estimated for a localized FRB \citep{cor03},
so one can define the extragalactic (or excess) dispersion measure of an FRB as
\begin{eqnarray}
{\rm DM_E}\equiv{\rm DM_{obs}}-{\rm DM_{MW}}={\rm DM_{IGM}}+{\rm DM_{HG}},
\label{dme}
\end{eqnarray}
which can be treated as an observed quantity.
The local DMs of FRB host galaxies may be assumed to have no significant evolution with redshift of $z\lesssim 1$, i.e. $\left\langle{\rm DM_{HG,loc}}\right\rangle\sim {\rm const}$, where $\left\langle{\rm DM_{HG,loc}}\right\rangle$ is the average value of the rest-frame host galaxy DM within a certain redshift bin. Due to cosmological time dilation, the observed host DM value reads ${\rm DM_{HG}} = {\rm DM_{HG,loc}} / (1+z)$ 
\citep{iok03}.
Considering the local inhomogeneity of the IGM \citep{mcq14}, we define a mean DM of the IGM as \citep{den14,yan16} 
\begin{eqnarray}
\langle{\rm{DM_{IGM}}}\rangle&=&\frac{3cH_0\Omega_bf_{\rm{IGM}}}{8\pi Gm_p} \int_0^z
\frac{f_e(z^\prime)(1+z^\prime)}
{\sqrt{\Omega_m(1+z^\prime)^3+\Omega_\Lambda}}dz^\prime,\nonumber\\
\nonumber\\
&\simeq&\frac{3cH_0\Omega_bf_{\rm{IGM}}f_e}{8\pi Gm_p}\left[z+z^2\left(\frac{1}{2}-\frac{3\Omega_m}{4}\right)+O(z^2)\right],\nonumber\\
\label{dm}
\label{dmigm}
\end{eqnarray}
where $f_e(z)=(3/4)y_1\chi_{e,\rm{H}}(z)+(1/8)y_2\chi_{e,\rm{He}}(z)$, $y_1\sim 1$ and $y_2\simeq 4-3y_1\sim 1$ are the hydrogen and helium mass fractions normalized to 3/4 and 1/4, respectively, and $\chi_{e,\rm{H}}(z)$ and $\chi_{e,\rm{He}}(z)$ are the ionization fractions for hydrogen and helium, respectively. For $z<3$, one has $\chi_{e,\rm{H}}(z)\simeq\chi_{e,\rm{He}}(z)\simeq1$, due to full ionization of both hydrogen and helium \citep{mei09}. Therefore, one has $f_e(z)\simeq f_e=7/8$. 
We adopt the flat ${\rm \Lambda CDM}$ parameters recently derived from the \emph{Planck} data: $H_0=67.7~\unit{km~s^{-1}Mpc^{-1}},~\Omega_m=0.31,~\Omega_\Lambda=0.69$, and $\Omega_b=0.049$ \citep{pla15}. For the fraction of baryon mass in the intergalactic medium, we adopt $f_{\rm{IGM}}=0.83$ \citep{fuk98,shu12}. In fact, due to IGM inhomogeneity, the fluctuation of individual measurements is expected \citep{mcq14}. Since the formation and evolution of the different galaxies are essentially independent at a given redshift $z$, ${\rm DM_{IGM}}(z)$ for different lines of sight would have a Gaussian distribution. As a result, the inhomogeneity of the IGM may affect the scatter, but not the mean trend of the DM-flux relation, especially when the sample size is large enough.

For an FRB with an intrinsic frequency-dependent isotropic-equivalent luminosity $L_{\nu_\ast}(\nu_\ast)$, the observed flux is given by $F_\nu d\nu=L_{\nu_\ast}d\nu_\ast/4\pi d_L^2$. The luminosity distance of the FRB may be given by
\begin{eqnarray}
d_L \simeq \left(\frac{L_{\rm iso}}{4\pi\nu F_\nu}\right)^{1/2},\label{flux}
\end{eqnarray}
where $L_{\rm iso}\equiv\nu_\ast L_{\nu_\ast}$ is the characteristic isotropic-equivalent luminosity (The true luminosity should be $L=(\Delta\Omega/4\pi)L_{\rm iso}$, where $\Delta\Omega$ is the beaming solid angle), and $\nu\simeq 1.4~\unit{GHz}$ is the characteristic frequency of FRBs\footnote{Strictly speaking, a proper $k$-correction is needed to derive a more rigorous $d_L$. However, the FRB spectral shape is not well constrained. Since the FRB emission seems to peak around 1 GHz and since the FRB redshift is not very high, our approximate treatment is justified.}.
For a flat universe, one has 
\begin{eqnarray}
d_L&=&\frac{c}{H_0}(1+z)\int_0^z
\frac{1}
{\sqrt{\Omega_m(1+z^\prime)^3+\Omega_\Lambda}}dz^\prime,
\nonumber\\
&\simeq&\frac{c}{H_0}\left[z+z^2\left(1-\frac{3\Omega_m}{4}\right)+O(z^2)\right].
\label{dl}
\end{eqnarray}
For $z\lesssim1$, according to Eq.(\ref{dme})-Eq.(\ref{dl}), we can obtain approximately ${\rm DM_{IGM}}\simeq A d_L$, where $A\equiv 3H_0^2\Omega_bf_{\rm{IGM}}f_e/8\pi Gm_p$. Therefore, one has the ${\rm DM_E}-F_\nu$ relation:
\begin{eqnarray}
\left\langle{\rm DM_E}\right\rangle\simeq \frac{A}{\sqrt{4\pi}} L_{\rm iso}^{1/2}\nu^{-1/2}F_\nu^{-1/2}+\left\langle{\rm DM_{HG,loc}}\right\rangle.\label{dmf}
\end{eqnarray}
As shown in Eq.(\ref{dmf}), $\left\langle{\rm DM_E}\right\rangle\propto F_\nu^{-1/2}$ for $F_\nu\ll F_{\nu,\rm{crit}}$, and $\left\langle{\rm DM_E}\right\rangle\simeq\left\langle{\rm DM_{HG,loc}}\right\rangle$ for $F_\nu\gg F_{\nu,\rm{crit}}$, where $F_{\nu,\rm{crit}}\equiv A^2L_{\rm iso}/4\pi\nu\left\langle{\rm DM_{HG,loc}}\right\rangle^2$.

One can numerically solve Eq.(\ref{dme})-Eq.(\ref{dl}), and use the observed ${\rm DM_E}-F_\nu$ relation to fit the current sample of 21 FRBs\footnote{The three bursts reported by \cite{cal17}, similar to the original ``lorimer'' burst \citep{lor07}, only have the lower limits of the peak fluxes reported.  In our analysis, these lower limits are used. Our conclusion of a large host galaxy DM remains valid if one adopts larger peak flux values for these bursts.}.
We take the FRB data from the {\em FRB Catalogue} of \cite{pet16}\footnote{
\url{http://www.astronomy.swin.edu.au/pulsar/frbcat/}} and ignore the effect of interstellar scintillation.
Except FRB 121102, other FRBs (mostly detected with Parkes) are not observed to repeat. If all FRB sources repeat, most bursts may be below the sensitivity of the Parkes telescope, and the detected one may be one of the brightest pulses.  For this reason, we take the brightest pulse of the repeater to define its peak flux at $1.4~\unit{GHz}$.\footnote{If one instead takes an average value of peak fluxes to denote the peak flux of FRB 121102, the inferred $\left\langle{\rm DM_{HG,loc}}\right\rangle$ from our analysis is even larger than reported, so our conclusion of a large host galaxy DM remains valid.}  

\begin{figure*}[t]
\centering
\includegraphics[angle=0,scale=0.45]{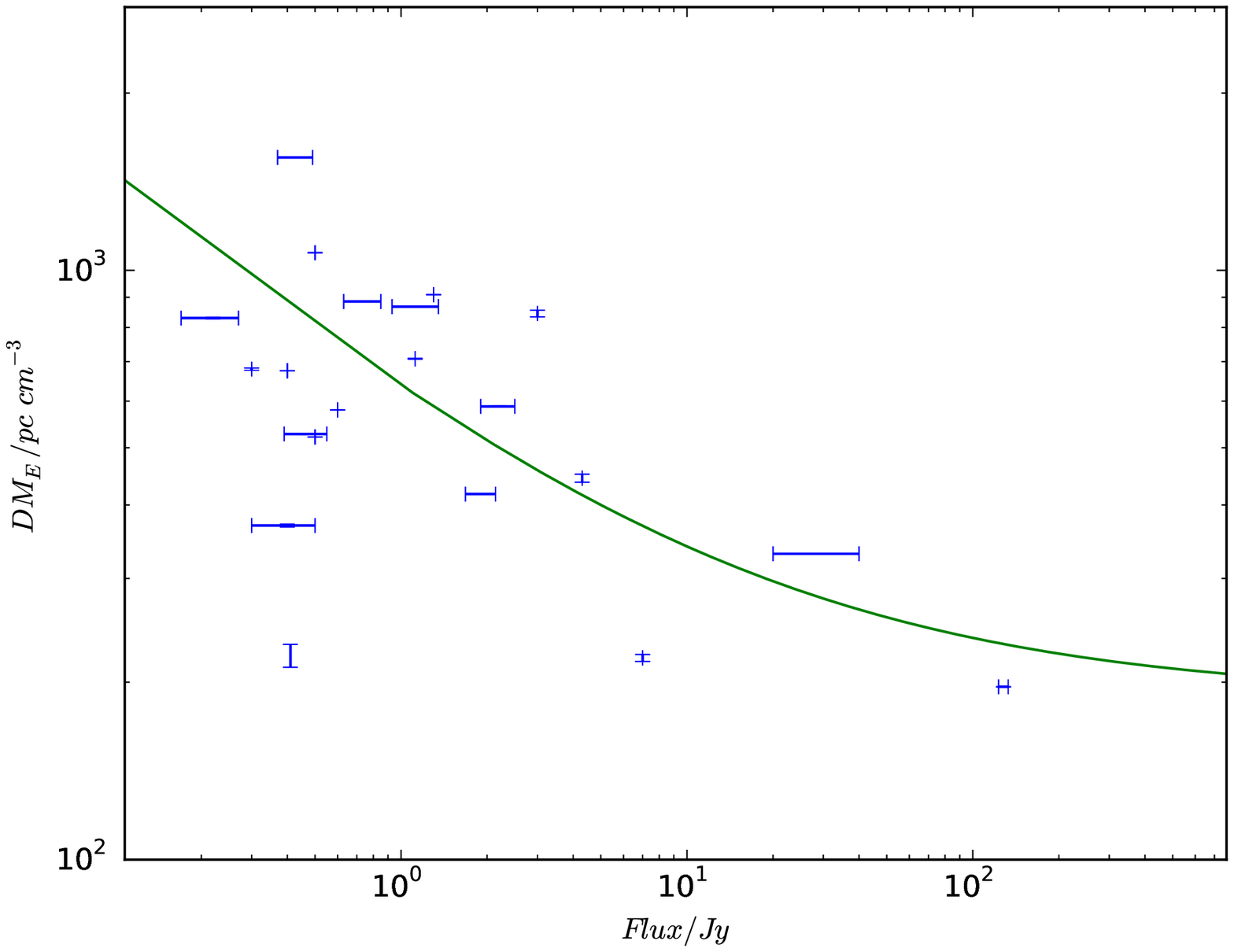}
\includegraphics[angle=0,scale=0.45]{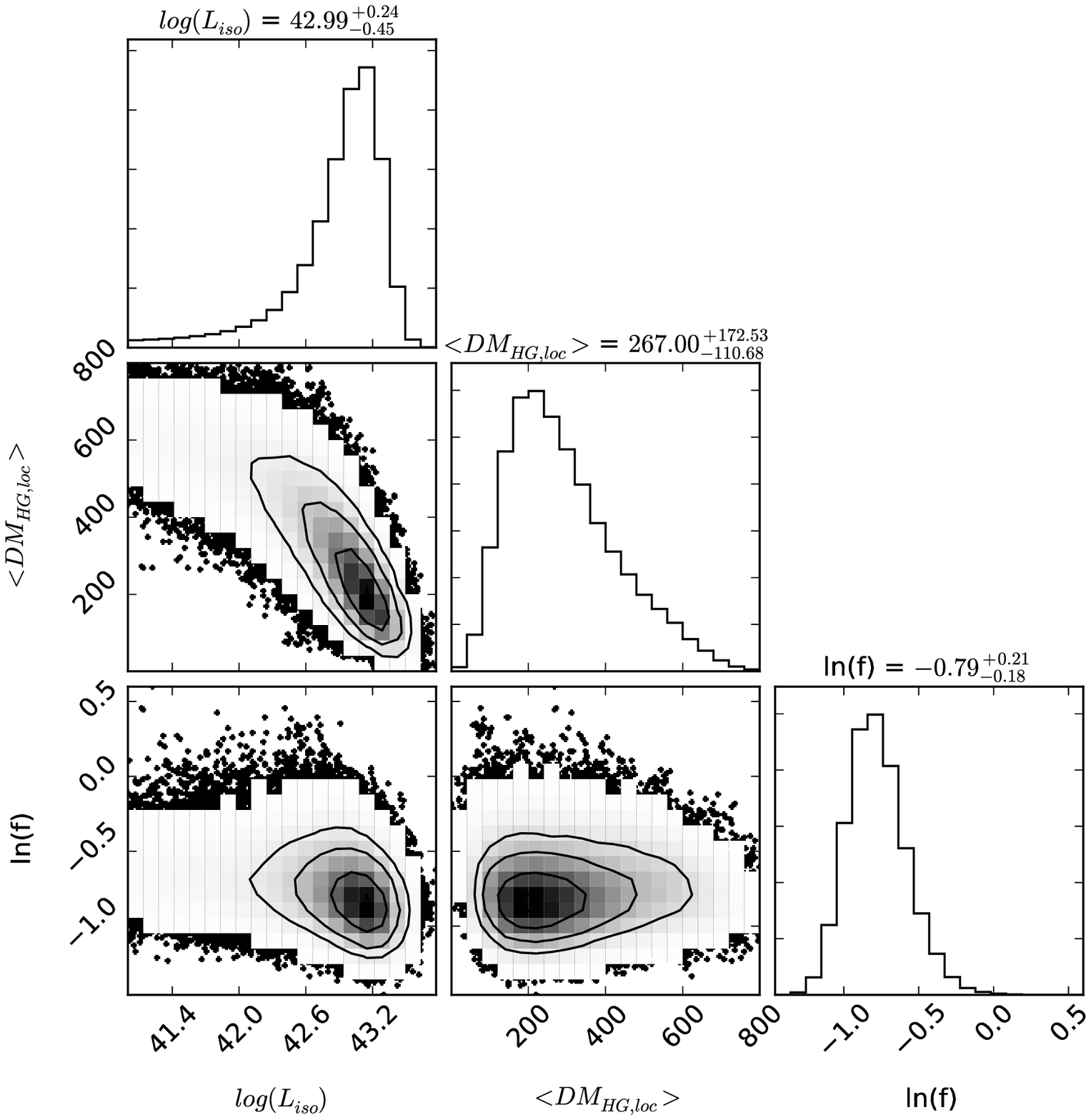}
\caption{Left: The best fit to the observed ${\rm DM_E} - F_\nu$ relation. The blue FRB data points are from the {\em FRB Catalogue}. The green line denotes the best fitting curve. Right: Two-dimensional projections of the posterior probability distributions of the model fitting parameters. Data points are shown as grayscale points with contours. Contours are shown at 0.5, 1, 1.5, and 2$\sigma$ significance levels. The best fitting values are shown on top of each 1D distribution. }\label{fig1}
\end{figure*}

We apply the Bayesian inference to extract $\left\langle{\rm DM_{HG,loc}}\right\rangle$ from the observed ${\rm DM_E}-F_\nu$ relation using the software \emph{emcee}\footnote{\url{http://dan.iel.fm/emcee/current.}}.
The log likelihood for the fitting parameters is determined by the $\chi^2$ statistics, i.e.
\begin{eqnarray}
\chi^2(L_{\rm iso},\langle{\rm DM_{HG,loc}}\rangle,f)
=\sum_{{i}}\frac{({\rm DM_{E,{\it i}}-\langle DM_E\rangle})^2}{\sigma_i^2+\sigma_{\rm sys}^2(f)},
\end{eqnarray}
where $i$ represents the sequence of an FRB in the sample, $\sigma_i$ represents the error of ${\rm DM_{E,i}}$, $\sigma_{\rm sys}\equiv f\langle{\rm DM_E}\rangle$ is the system error, and $f$ is a fitting parameter reflecting the uncertainty of the model. 
At first, we use the uninformative priors on $\log(L_{\rm iso})$, $\left\langle{\rm DM_{HG,loc}}\right\rangle$ and $\ln f$. Combining the priors with the definition of log likelihood from above, one can obtain the log-probability function. Then we initialize the walkers in a tiny Gaussian ball around the maximum likelihood result and sample the probability distribution. We then get the projections of the posterior probability distributions of the model fitting parameters in Figure \ref{fig1}.

The analysis results are shown in Figure \ref{fig1}. We have $\log(L_{\rm iso}/\unit{erg~s^{-1}})=42.99_{-0.45}^{+0.24}$ (under the assumption of a constant $L_{\rm iso}$), $\left\langle{\rm DM_{HG,loc}}\right\rangle=267.00_{-110.68}^{+172.53}~\unit{pc~cm^{-3}}$ and $\ln f=-0.79_{-0.18}^{+0.21}$. Our results show that FRBs may have a large host galaxy DM, although with a large dispersion. 
To apply this method, we have to assume that the isotropic-equivalent luminosity of the FRBs has a characteristic value $L_{\rm iso}$, which requires that $L_{\rm iso}$ has a narrow distribution $\Phi(L_{\rm iso})$. We perform a series of Monte Carlo simulations to test how narrow the isotropic luminosity function needs to be in order to correctly derive the prior $\left\langle{\rm DM_{HG,loc}}\right\rangle$ with $\sim1\sigma$ accuracy. We find that for a power-law distribution  $\Phi(L_{\rm iso})\propto L_{\rm iso}^{-\alpha}$, one needs to have $\alpha>3$; for a Gaussian distribution, one needs to have $\Phi(L_{\rm iso})\propto N(L_{\rm iso,mean},<0.3L_{\rm iso,mean})$.

Our derived relatively large value of $\left\langle{\rm DM_{HG,loc}}\right\rangle$ is supported by two other independent pieces of evidence.

First, for Galactic pulsars, the scattering time is found to be $\tau\ll10^{-3}~\unit{ms}$ for $|b|\gtrsim30^\circ$ \citep[e.g.][]{cor16b}. However, some FRBs with $|b|>30^\circ$ have measured scattering time of a few milliseconds \citep{pet16}, suggesting that scattering happens outside the Milky Way. The scattering contribution from the IGM is calculated to be negligibly small, so that most the scattering would occur in the FRB host galaxy \citep{lua14,xu16}. Observations show that a larger scattering tail correspond to a larger DM, e.g. $\hat \tau=2.98\times10^{-7}~\unit{ms}~{\rm DM}^{1.4}(1+3.55\times10^{-5}{\rm DM}^{3.1})$ for Galactic pulsars \citep{cor16b}, a fact understandable with the turbulence theories \citep{xu17}. 
If one assumes that the FRBs' hosts have a similar relation, 
for the millisecond scattering time, one would require ${\rm DM_{HG}}\gtrsim280~\unit{pc~cm^{-3}}$.
This is consistent with our derived results.

Second, the host-galaxy DM of FRB 121102 may be inferred based on the data. 
FRB 121102 was localized to a $\sim0.1$ arcsecond precision by \citet{cha17}. The observed dispersion measure is ${\rm DM_{obs}}=558~\unit{pc~cm^{-3}}$. According to the location of FRB 121102, \citet{ten17} identified an extended source coincident with the burst, which is a host galaxy at $z=0.19273$. Adopting the {\em Planck} cosmological parameters and $f_{\rm IGM}=0.83$, one derives ${\rm{DM_{IGM}}}\simeq164~\unit{pc~cm^{-3}}$ (subject to local fluctuations \citep{mcq14}). The MW contribution is ${\rm DM_{MW}}\simeq218~\unit{pc~cm^{-3}}$ in the direction \citep{cha17}. So one may derive 
\begin{equation}
{\rm DM_{HG}}={\rm DM_{obs}}-{\rm DM_{MW}}-{\rm DM_{IGM}}\simeq176~\unit{pc~cm^{-3}}
\end{equation}
and 
\begin{equation}
{\rm DM_{HG,loc}}= (1+z) {\rm DM_{HG}} \simeq 210~\unit{pc~cm^{-3}}
\end{equation}
for FRB 121102. Such a value is also consistent with our fitting results and previous constraints \citep{ten17}.

\section{Discussion}

The above results suggest that the FRB host galaxies have a relatively large value of DM. There  could be two possible contributions to such a large DM: the ISM in the host galaxy and the near-source plasma. 

For the case of a host ISM, one immediate inference is that the type of the host galaxies of most FRBs would not be MW-like disk galaxies. The reason is that for disk galaxies, FRBs would be most likely emitted from high galactic latitudes, which gives rise to negligible DM values\footnote{If FRB sources are associated with the center of galaxies, DMs from disk galaxies could be large in general.}. Indeed, the host galaxy of FRB 121102 was identified as a dwarf galaxy \citep{ten17}, which is consistent with our expectation. However, our inferred value is still somewhat larger than the simulated host galaxy DM for various types of galaxies \citep{xu15}, suggesting that a near-source plasma may be needed.

For the case of a near-source plasma, we consider the contributions from a SNR, a PWN, and an HII region. First,
in a thin shell approximation, the DM value through a young SNR may be estimated by \citep[see also][]{kat16,pir16,pir17,met17}
\begin{eqnarray}
{\rm DM_{SNR}}&=&\frac{M}{4\pi \mu_m m_p R^2}\nonumber\\
&=&272~\unit{pc~cm^{-3}}\frac{M}{M_\odot}\left(\frac{R}{0.1~\unit{pc}}\right)^{-2},
\end{eqnarray}
where $M$ and $R$ are the SNR mass and radius, respectively, and $\mu_m=1.2$ is the mean molecular weight. Note that the SNR dispersion measure does not depend on the thickness of the thin shell. The DM variation of the SNR during an observation time $\Delta t$ is given by
\begin{eqnarray}
\Delta {\rm DM_{SNR}}&=&\frac{M \upsilon}{2\pi\mu_m m_p R^3}\Delta t\nonumber\\
&=&16.7~\unit{pc~cm^{-3}}\frac{M}{M_\odot}\left(\frac{R}{0.1~\unit{pc}}\right)^{-3}\nonumber\\
&\times&\left(\frac{\upsilon}{3000~\unit{km~s^{-1}}}\right)\left(\frac{\Delta t}{1~\unit{yr}}\right),
\end{eqnarray}
where $\upsilon\sim(3000-30000)~\unit{km~s^{-1}}$ is the characterized SNR velocity. The age of the SNR may be estimated as $T\simeq R/\upsilon\simeq(3-30)~\unit{yr}(R/0.1~\unit{pc})$.
In principle, it is possible to expect that the host DM is dominated by a supernova ejecta. However, there are two caveats for this possibility:  
1. The thin shell model predicts a secular variation in DM, which needs to be confirmed by long-term observations.
2. For an age-independent event rate (the time delay between SN and FRB is uniformly distributed for FRBs), the cumulative distribution of DMs of FRBs should satisfy $N(>{\rm DM})\propto {\rm DM}^{-1/2}$ if the observed DM is dominated by the SNRs associated with the FRBs. However, the statistical results of the observed FRBs obviously deviate from this relation \citep{kat16}.

Next, we consider the DM contribution from a PWN. 
Some authors suggested an association of FRBs with young pulsars \citep{cor16,con16}, while some others suggested an association of FRBs with magnetar giant flares \citep{pop10,kul14}. While these models are greatly constrained by available observations \citep{ten16,lyu17}, we nonetheless consider the DM contribution from a pulsar/magnetar wind. 
A relativistic electron-positron pair plasma is expected to stream out from the magnetosphere, and the number density of the wind at radius $r$ is given by \citep[e.g.][]{mur16a,cao17,dai17}
\begin{eqnarray}
n_{\rm w}(r)\simeq\frac{\dot N_{\rm w}}{4\pi r^2c}=\mu_\pm n_{\rm GJ}(R_{\rm LC})\left(\frac{r}{R_{\rm LC}}\right)^{-2},
\end{eqnarray}
where $R_{\rm LC}$ is the radius of the light cylinder, $\mu_\pm$ is pair multiplicity parameter, $n_{\rm GJ}=(\Omega B_p/2\pi ec)(r/R)^{-3}$ is the classical Goldreich-Julian number density \citep{gol69}, $\dot N_{\rm w}\simeq 4\pi R_{\rm LC}^2\mu_\pm n_{\rm GJ}(R_{\rm LC})c$ is the particle number flux, $B_p$ is the polar-cap magnetic field strength, and $R$ is the neutron star radius. The DM of the pulsar/magnetar wind is given by \citep[e.g.][]{cao17}
\begin{eqnarray}
{\rm DM_w}&\simeq&\int_{R_{\rm LC}}^{R_{\rm sh}} 2\Gamma(r)n_w(r)dr\nonumber\\
&\simeq&3\Gamma_L\mu_\pm n_{\rm GJ}(R_{\rm LC})R_{\rm LC}\left[1-\left(\frac{R_{\rm sh}}{R_{\rm LC}}\right)^{-2/3}\right]\nonumber\\
&\simeq&146~\unit{pc~cm^{-3}}\left(\frac{\mu}{10^6}\right)^{2/3}\left(\frac{B_p}{10^{14}~\unit{G}}\right)^{4/3}\nonumber\\
&\times&\left(\frac{P}{0.3~\unit{s}}\right)^{-11/3}~~~(R_{\rm sh}\gg R_{\rm LC}),
\end{eqnarray}
where $R_{\rm sh}$ is the radius of the shock, $P$ is the rotation period, $\Gamma_L\sim (L_{\rm sd}/\dot N_{\rm w}m_ec^2)^{1/3}$ is the relativistic wind Lorentz factor at the light cylinder, and $L_{\rm sd}=B_p^2R^6\Omega^4/6c^3$ is the pulsar/magnetar spin-down luminosity. The wind Lorentz factor at radius $r$ may be given by $\Gamma(r)\sim\Gamma_L(r/R_{\rm LC})^{1/3}$. On the other hand, in the shock, electron-positron pairs are thermalized. They would undergo cooling and may become non-relativistic. For the PWN with its age much longer than the spindown time $T_{\rm sd}$, the dispersion measure from these thermalized particles is given by 
\begin{eqnarray}
{\rm DM_{sh}}&\simeq&\frac{\dot N_{\rm w}T_{\rm sd}}{4\pi R_{\rm sh}^2}=\frac{3c^2\mu_\pm I}{2\pi eB_pR^3R_{\rm sh}^2}\nonumber\\
&\simeq&3\times10^{-5}\unit{pc~cm^{-3}}\left(\frac{\mu_\pm}{10^6}\right)\left(\frac{B_p}{10^{14}~\unit{G}}\right)^{-1}\nonumber\\
&\times&\left(\frac{R_{\rm sh}}{0.1~\unit{pc}}\right)^{-2},
\end{eqnarray}
where $I\simeq 10^{45}~\unit{g~cm^2}$ is the moment of inertia, $\dot N_{\rm w}T_{\rm sd}$ is the electron-positron pair number ejected over the spindown time $T_{\rm sd}$, which does not depend on $\Omega$.
Notice that the DM contribution from the thermalized pairs in the shock could be ignored. Therefore, the total DM from PWN is ${\rm DM_{PWN}}={\rm DM_w}+{\rm DM_{sh}}\simeq{\rm DM_w}$. The pulsar/magnetar wind may provide a significant contribution to DM if $\mu_\pm$ is large enough.

Recently, \citet{zha17} proposed a unified interpretation of FRBs in the so-called ``cosmic comb'' model, which invokes the interaction between an astrophysical plasma stream and a foreground regular pulsar. Since cosmic combs more easily happen in slow ($P\sim1~\unit{s}$) and low-field ($B\sim10^{12}$~\unit{G}) pulsars, the DM contribution from the near-source plasma is ${\rm DM_{PWN}}\sim0.003~\unit{pc~cm^{-3}}$. Therefore, in the cosmic comb model, the large host galaxy DM might result from the host galaxy ISM or the near-source plasma of the stream source in front of the pulsar towards Earth.

At last, we consider the DM contribution from a HII region in the host galaxy, assuming that an FRB is embedded in a Str\"omgren sphere. The DM contributed by a Str\"omgren sphere may be estimated as
\begin{eqnarray}
{\rm DM_{HII}}&\simeq&nR_{\rm str}=\left(\frac{3N_un}{4\pi\alpha_{\rm B}}\right)^{1/3}\nonumber\\
&=&540~\unit{pc~cm^{-3}}\left(\frac{N_u}{5\times10^{49}~\unit{s}^{-1}}\right)^{1/3}\left(\frac{n}{100~\unit{cm^{-3}}}\right)^{1/3},\nonumber\\
\end{eqnarray}
where $n$ is the gas number density in the HII region, $N_u$ is the rate of ionizing photons from a star, $\alpha_B$ is the recombination rate, and $R_s\equiv\left(3N_u/4\pi\alpha_{\rm B} n^2\right)^{1/3}$ is the Str\"omgren radius. We assume that there is an O5 star in the HII region, so that $\alpha_{\rm B}=2.6\times10^{-13}~\unit{cm^3s^{-1}}$ for $T=10^{4}~\unit{K}$, and the Str\"omgren radius is $R_{\rm str}=5.4~\unit{pc}$. We note that the Str\"omgren radius is much larger than the projected size of $\lesssim0.7~\unit{pc}$ of FRB 121102 radio persistent emission source \citep{mar17}. 

An FRB may be absorbed by the HII region via free-free absorption.
In the Rayleigh-Jeans limit, the free-free absorption coefficient is given by \citep[e.g.][]{lua14}
\begin{eqnarray}
\alpha_{\rm ff}&=&\frac{4}{3}\left(\frac{2\pi}{3}\right)^{1/2}\frac{Z^2e^6n_en_i\bar g_{\rm ff}}{c m_e^{3/2}(k_BT)^{3/2}\nu^2},\nonumber\\
\bar g_{\rm ff}&=&\frac{\sqrt 3}{\pi}\left[\ln\left(\frac{(2k_BT)^{3/2}}{\pi e^2m_e^{1/2}\nu}\right)-\frac{5}{2}\gamma\right],
\end{eqnarray}
where $n_e$ and $n_i$ are the number densities of electrons and ions, respectively, $\gamma=0.577$ is Euler's constant and $\bar g_{\rm ff}$ is the Gaunt factor. For an HII region, one may assume $n_e=n_i$ and $Z=1$. The optical depth for the free-free absorption is $\tau\sim\alpha_{\rm ff} R_{\rm str}$, which gives
\begin{eqnarray}
\tau\simeq0.018\left(\frac{n}{100~\unit{cm^{-3}}}\right)^{4/3}\left(\frac{T}{10^4~\unit{K}}\right)^{-1.5}\left(\frac{\nu}{1~\unit{GHz}}\right)^{-2},\nonumber\\
\end{eqnarray} 
where we have taken $\alpha_{\rm B}=2.6\times10^{-13}~\unit{cm^3~s^{-1}}$ and $\bar g_{\rm ff}=6.0$ for $T=10^4~\unit{K}$ and $\nu=1~\unit{GHz}$. Therefore, such a HII region is optically thin for FRBs.

In summary, we show that the current FRB observations imply large host galaxy DM values, e.g., $\left\langle{\rm DM_{HG,loc}}\right\rangle\gtrsim200~\unit{pc~cm^{-3}}$. Such a large DM may be contributed by the host ISM or a near-source plasma.
Such a result poses requirements to FRB progenitor models. 
 
For the models invoking young energetic pulsars and magnetars \citep[e.g.][]{con16,cor16,yan16a,pir16,mur16a,mur16b,met17,kas17,dai17} or collapse of new-born supra-massive neutron star \citep[e.g.][]{fal14,zha14}, a near-source SNR, PWN or HII region would give an important contribution to the observed DM. Also irregular star-forming galaxies (e.g. the host galaxy of FRB 121102) do not have a disk-like structure, and would provide a relatively large host DM. So these models are more consistent with the large host DM inferred from this paper.  For the models invoking compact object mergers \citep[e.g.][]{tot13,zha16,wan16}, the contribution from a near-source plasma may not be important (except for ``prompted'' mergers that have a short time delay from star formation). Some of these systems may also have a large offset from the host galaxy, which may not give a large local DM. However, since mergers can happen in elliptical or early-type host galaxies, a relatively large  ${\rm DM_{HG,loc}}$ may arise from a large free electron column from the extended halo of these galaxies. 

In our analysis we ignored the effects of interstellar scintillation and host galaxy evolution. Interstellar scintillation, if significant, may affect the detectability of FRBs \citep{cor17}. In our analysis, we introduced one parameter to denote the instrumental systematic errors in our simulations. This factor may partially account for the uncertainty of FRB flux introduced from interstellar scintillation. The evolution of the FRB host galaxy might lead to the DM evolution of the ISM component in the host galaxy, which depends on the host-galaxy morphology, mass, and star formation. However, if ${\rm DM_{HG,loc}}$ is dominated by the contribution from the near-source plasma, the cosmological evolution effect of the host galaxies may be smeared.  

\acknowledgments
We thank the anonymous referee for detailed suggestions that have allowed us to improve this manuscript significantly. We also thank Keith Bannister, Zi-Gao Dai, Tian-Qi Huang, Yan Huang, Shriharsh Tendulkar, Su Yao, and Hai Yu for helpful discussions and comments. This work is partially supported by the Initiative Postdocs Supporting Program (No. BX201600003), the National Basic Research Program (973 Program) of China (No. 2014CB845800), the National Natural Science Found (No.11273005) and Project funded by China Postdoctoral Science Foundation (No. 2016M600851). Y.-P.Y. is supported by a KIAA-CAS Fellowship.

\end{document}